\documentclass[epj,twocolumn]{webofc}
\usepackage[varg]{txfonts}   
\woctitle{TRANSVERSITY 2014}
\begin{document}
\title{Dihadron Fragmentation Functions and Transversity}
%
%

\author{Marco Radici\inst{1}\fnsep\thanks{\email{marco.radici@pv.infn.it}} \and
        A. Courtoy\inst{2}\fnsep\thanks{\email{aurore.courtoy@ulg.ac.be}} \and
        Alessandro Bacchetta\inst{3,1}\fnsep\thanks{\email{alessandro.bacchetta@unipv.it}}
}
    
\institute{INFN - Sezione di Pavia, via Bassi 6, I-27100 Pavia, Italy
\and
           IFPA, AGO Department, Universit\'e de Li\`ege, Bat. B5, Sart Tilman B-4000 Li\`ege, Belgium
\and
           Dipartimento di Fisica, Universit\`a di Pavia, via Bassi 6, I-27100 Pavia, Italy     
     }

\abstract{We present preliminary results for an updated extraction of the transversity parton distribution based on the analysis of pion-pair production in deep-inelastic scattering off transversely polarized targets in collinear factorization. Data for proton and deuteron targets by HERMES and COMPASS allow for a flavor separation of the valence components of transversity, while di-hadron fragmentation functions are taken from the semi-inclusive production of two pion pairs in back-to-back jets in $e^+ e^-$ annihilation. The latter data from Belle have been reanalyzed using the replica method and a more realistic estimate of the uncertainties on the chiral-odd interference fragmentation function has been obtained. After encoding this piece of information into the deep-inelastic scattering cross section, the transversity has been re-extracted by using the most recent and more precise COMPASS data for proton target. This picture represents the current most realistic estimate of the uncertainties on our knowledge of transversity. The preliminary results indicate that the valence up component seems smaller and with a narrower error band than in previous extraction.}

\maketitle
\section{Introduction}
\label{sec:intro}

Parton distribution functions (PDFs) describe combinations of number densities of quarks and gluons in a fast-moving hadron.  At leading twist, the spin structure of spin-half hadrons is specified by three PDFs. The least known one is the transverse polarization (transversity) distribution $h_1$ because, being chiral-odd, it can be measured only in processes with two hadrons in the initial state, or one hadron in the initial state and at least one hadron in the final state ({\it e.g.} semi-inclusive DIS - SIDIS).  

Combining data on polarized single-hadron SIDIS together with data on almost back-to-back emission of two hadrons in $e^+ e^-$ annihilations, the transversity distribution was extracted for the first time by the Torino 
group~\cite{Anselmino:2008jk,Anselmino:2013vqa}. The weak part of this analysis lies in the factorization framework used to interpret the data, since it involves Transverse Momentum Dependent partonic functions (TMDs). QCD evolution of TMDs must be included to analyze SIDIS and $e^+ e^-$ data obtained at very different scales, but an active debate is ongoing about which is the correct framework to compute these effects~\cite{Collins:2014loa}.  

Alternatively, transversity can be extracted in the standard framework of collinear factorization using SIDIS with two hadrons detected in the final state. In fact, $h_1$ is multiplied with a specific chiral-odd Di-hadron Fragmentation Function 
(DiFF)~\cite{Collins:1994kq,Jaffe:1998hf,Radici:2001na}, which can be extracted from the corresponding $e^+ e^-$ annihilation process leading to two back-to-back pion pairs~\cite{Boer:2003ya,Courtoy:2012ry}. In the collinear framework, evolution equations of DiFFs can be easily computed~\cite{Ceccopieri:2007ip}. Using two-hadron SIDIS data on proton and deuteron targets from HERMES~\cite{Airapetian:2008sk} and COMPASS~\cite{Adolph:2012nw}, as well as Belle data for the process 
$e^+ e^- \to (\pi^+ \pi^-) (\pi^+ \pi^-) X$~\cite{Vossen:2011fk}, the transversity $h_1$ was extracted for the first time in the collinear framework~\cite{Bacchetta:2011ip} and the valence components of up and down quark were 
separated~\cite{Bacchetta:2012ty}. 

In this contribution, we update the extraction of DiFFs from $e^+ e^-$ annihilation data by re-performing the fit using the replica method~\cite{Bacchetta:2012ty}. Then, using the most recent SIDIS data for unidentified hadron pairs on a proton target by COMPASS~\cite{Adolph:2014fjw} we re-extract the $h_1$, thus getting the current most realistic estimate of the uncertainties on our knowledge of it.

\section{Theoretical Framework for Di-hadron Semi-Inclusive Production}
\label{sec:theory}

We consider the process $\ell(k) + N(P) \to \ell(k') + H_1(P_1) + H_2(P_2) + X$, where $\ell$ denotes the beam lepton, $N$ the nucleon target with mass $M$ and polarization $S$, $H_1$ and $H_2$ the produced unpolarized hadrons with masses $M_1$ and $M_2$, respectively. We define the total $P_h = P_1 + P_2$ and relative $R = (P_1-P_2)/2$ momenta of the pair, with $P_h^2 = M_h^2 \ll Q^2=-q^2$ and $q = k - k'$ the momentum transferred. We define the azimuthal angles $\phi_R$ and 
$\phi_S$ as the angles of ${\bf R}_T$ and ${\bf S}_T$, respectively, around the virtual photon direction ${\bf q}$. We also define the polar angle $\theta$ between the direction of the back-to-back emission in the center-of-mass (cm) frame of the two hadrons and the direction of $P_h$ in the photon-proton cm frame. Then, ${\bf R}_T = {\bf R} \sin\theta$ and $|{\bf R}|$ is a function of the invariant mass only~\cite{Bacchetta:2002ux}. Finally, we use the standard definition of the SIDIS invariants $x,\, y$ for the fractional momentum carried by quarks and for the beam fractional energy delivered to them, respectively; the fractional energy carried by the final hadron pair is $z = z_1 + z_2$. To leading twist, the differential cross section for the two-hadron SIDIS off a transversely polarized nucleon target becomes~\cite{Bacchetta:2012ty}
\begin{eqnarray}
& &\frac{d\sigma}{dx \, dy\, dz\, d\phi_S\, d\phi_R\, d M_{h}^2\,d \cos{\theta}} =  \frac{\alpha^2}{x y\, Q^2}  \nonumber \\
& &\times \Biggl\{ A(y) \, F_{UU} + |\boldmath{S}_T|\, B(y) \, \sin(\phi_R+\phi_S)\,  F_{UT} \Biggr\} 
\label{eq:crossSIDIS}
\end{eqnarray}
where $\alpha$ is the fine structure constant, $A(y) = 1-y+y^2/2$, $B(y) = 1-y$, and 
\begin{eqnarray} 
F_{UU} & = &x \sum_q e_q^2\, f_1^q(x; Q^2)\, D_1^q\bigl(z,\cos \theta, M_h; Q^2\bigr) \; , \nonumber \\
F_{UT} &=  &\frac{|{\bf R}| \sin \theta}{M_h}\, x \nonumber \\
& &\times \sum_q e_q^2\,  h_1^q(x; Q^2)\,H_1^{\sphericalangle\, q}\bigl(z,\cos \theta, M_h; Q^2\bigr) \; , \nonumber \\
& &   \label{eq:StructFunct}
\end{eqnarray}
with $e_q$ the fractional charge of a parton with flavor $q$. The $D_1^q$ is the DiFF describing the hadronization of an unpolarized parton with flavor $q$ into an unpolarized hadron pair. The $H_1^{\sphericalangle\, q}$ is its 
chiral-odd partner describing the same fragmentation but for a transversely polarized parton~\cite{Bianconi:1999cd}. DiFFs can be expanded in Legendre polynomials in $\cos \theta$~\cite{Bacchetta:2002ux}. After averaging over $\cos \theta$, only the term corresponding to the unpolarized pair being created in a relative $\Delta L=0$ state survives in the $D_1$ expansion, while the interference in $|\Delta L| = 1$ survives for $H_1^{\sphericalangle}$~\cite{Bacchetta:2002ux}. Without ambiguity, the two terms will be identified with $D_1$ and $H_1^{\sphericalangle}$, respectively. 

Inserting the structure functions of Eq.~(\ref{eq:StructFunct}) into the cross section~(\ref{eq:crossSIDIS}), we get the single-spin asymmetry (SSA)~\cite{Radici:2001na,Bacchetta:2002ux,Bacchetta:2006un}
\begin{eqnarray}
& &A_{{\rm SIDIS}}(x, z, M_h; Q) =  - \frac{B(y)}{A(y)} \,\frac{|\boldmath{R} |}{M_h} \nonumber \\
& &\qquad \times \frac{ \sum_q\, e_q^2\, h_1^q(x; Q^2)\, H_1^{\sphericalangle\, q}(z, M_h; Q^2)    } 
                                       { \sum_q\, e_q^2\, f_1^q(x; Q^2)\, D_{1}^q (z, M_h; Q^2) }\;  .
\label{eq:SIDISssa}
\end{eqnarray} 
For the specific case of $\pi^+ \pi^-$ production, isospin symmetry and charge conjugation suggest $D_1^q = D_1^{\bar{q}}$ and $H_1^{\sphericalangle\, q} = - H_1^{\sphericalangle\, \bar{q}}$, with $q=u,d,s$, with also 
$H_1^{\sphericalangle\, u} = - H_1^{\sphericalangle\, d}$~\cite{Bacchetta:2006un,Bacchetta:2011ip,Bacchetta:2012ty}. Moreover, from Eq.~(\ref{eq:SIDISssa}) the $x$-dependence of transversity is more conveniently studied by integrating the $z$- and $M_h$-dependences of DiFFs. So, the actual combinations used in the SIDIS analysis are, for the proton 
target~\cite{Bacchetta:2012ty}, 
\begin{eqnarray} 
x\, h_1^{p}(x; Q^2) &\equiv &x \, h_1^{u_v}(x; Q^2) - {\textstyle \frac{1}{4}}\, x h_1^{d_v}(x; Q^2) \nonumber \\
&= &-\frac{ A^p_{{\rm SIDIS}} (x; Q^2)  }{n_u^{\uparrow}(Q^2)}\,\frac{A(y)}{B(y)} \, \frac{9}{4} \nonumber \\
& &\times \sum_{q=u,d,s} \, e_q^2\, n_q (Q^2)\, x f_1^{q+\bar{q}}(x; Q^2) \; , 
\label{eq:xh1p}
\end{eqnarray}
and, for the deuteron target, 
\begin{eqnarray} 
 x\, h_1^{D} (x; Q^2) &\equiv &x \, h_1^{u_v}(x; Q^2)+ x h_1^{d_v}(x; Q^2)   \nonumber \\
 &= &- \frac{A^D_{\text{SIDIS}}(x; Q^2)}{n_u^{\uparrow}(Q^2)} \,\frac{A(y)}{B(y)} \, 3 \nonumber \\
 & &\times \sum_{q=u,d,s}\, 
 \big[ e_q^2\, n_q (Q^2) + e_{\tilde{q}}^2\, n_{\tilde{q}} (Q^2) \big] \nonumber \\
 & &\hspace{1.5cm} \times \; x f_1^{q+\bar{q}}(x; Q^2) \; , 
\label{eq:xh1D}
\end{eqnarray}
where $h_1^{q_v} \equiv h_1^q - h_1^{\bar{q}}$, $f_1^{q+\bar{q}} \equiv f_1^q + f_1^{\bar{q}}$, $\tilde{q}=d,u,s$ if $q=u,d,s$, respectively ({\it i.e.} it reflects isospin symmetry of strong interactions inside the deuteron), and 
\begin{eqnarray} 
n_q(Q^2) &= &\int  \int dz \, dM_h \, D_1^q (z, M_h; Q^2)  \nonumber  \\
n_q^\uparrow (Q^2) &= &\int \int dz \, dM_h \, \frac{|{\bf R}|}{M_h}\, H_1^{\sphericalangle\, q}(z,M_h; Q^2) \; .
\label{eq:DiFFnq}
\end{eqnarray}

\section{Extraction of Di-hadron Fragmentation Functions}
\label{sec:DiFFs}

The quantities in Eq.~(\ref{eq:DiFFnq}) can be determined by extracting DiFFs from the $e^+ e^- \to (\pi^+ \pi^-) (\pi^+ \pi^-) X$ process. In fact, the leading-twist cross section in collinear factorization, namely by integrating upon all transverse momenta but ${\bf R}_T$ and ${\bf \bar{R}}_T$, can be written as~\cite{Courtoy:2012ry}
\begin{equation}
d\sigma = \frac{1}{4\pi^2}\, d\sigma^0 \, \bigg( 1+ \cos (\phi_R + \phi_{\bar{R}} ) \, A_{e+e-} \bigg) \; , 
\label{eq:e+e-cross}
\end{equation}
where the azimuthal angles $\phi_R$ and $\phi_{\bar{R}}$ give the orientation of the planes containing the momenta of the pion pairs with respect to the lepton plane (see Fig.1 of Ref.~\cite{Courtoy:2012ry} for more details), and we define the so-called 
Artru-Collins asymmetry~\cite{Boer:2003ya}
\begin{eqnarray}
& &A_{e+e-} \propto \frac{|{\bf R}_T|}{M_h} \, \frac{|{\bf \bar{R}}_T|}{\bar{M}_h} \nonumber \\
& &\quad \times \frac{\sum_q e_q^2\, H_1^{\sphericalangle\, q}(z,M_h; Q^2)\, H_1^{\sphericalangle\, \bar{q}}(\bar{z},\bar{M}_h; Q^2)}
        {\sum_q e_q^2\, D_1^q(z,M_h; Q^2)\, D_1^{\bar{q}}(\bar{z},\bar{M}_h; Q^2)} \, .
\label{eq:e+e-ssa}
\end{eqnarray} 

Since a measurement of the unpolarized $e^+ e^-$ cross section is still missing, the unpolarized DiFF $D_1$ was parametrized to reproduce the two-pion yield of the PYTHIA event generator tuned to the Belle kinematics~\cite{Courtoy:2012ry}. The fitting expression at the starting scale $Q_0^2=1$ GeV$^2$ was inspired by previous model 
calculations~\cite{Bacchetta:2006un,Radici:2001na,Bianconi:1999uc,Bacchetta:2008wb} and it contains three resonant channels (pion pair produced by $\rho$, $\omega$, and $K^0_S$ decays) and a continuum. For each channel and for each flavor $q= u,d,s,c$, a grid of data in $(z,M_h)$ was produced using PYTHIA for a total amount of approximately 32000 bins. Each grid was separately fitted using the corresponding parametrization of $D_1$ and evolving it to the Belle scale at $Q^2=100$ GeV$^2$. An average $\chi^2$ per degree of freedom ($\chi^2$/dof) of 1.62 was reached using in total 79 parameters (see Ref.~\cite{Courtoy:2012ry} for more details).  

\begin{figure}
\centering
\includegraphics[width=7cm]{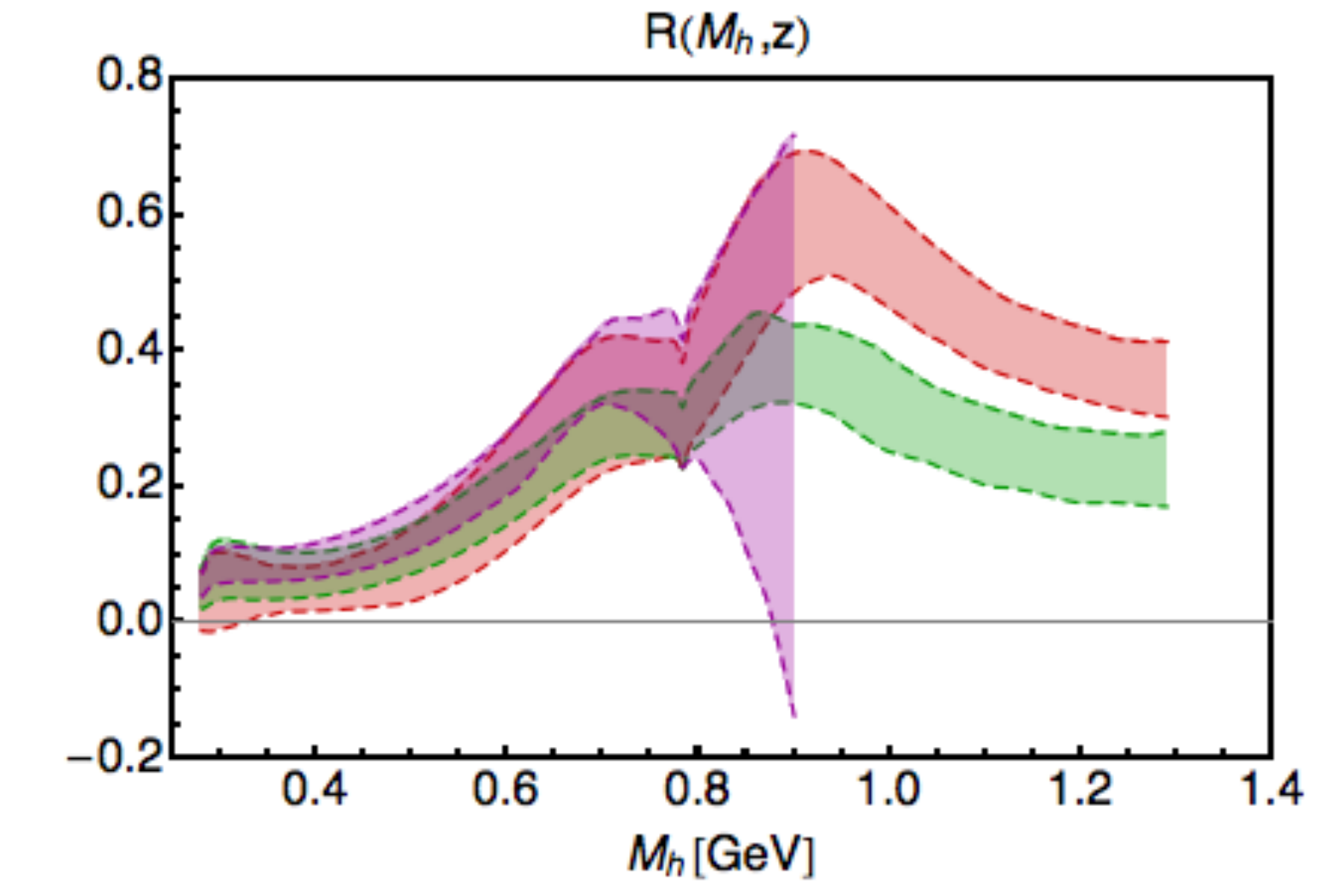}
\caption{$(|{\bf R}| / M_h) \, (H_1^{\sphericalangle\, u} / D_1^u)$ as a function of $M_h$ at $Q_0^2=1$ GeV$^2$ for three different $z=0.25$ (shortest violet band), $z=0.45$ (lower green band), and $z=0.65$ (upper red band).} 
\label{fig:H1Mh}
\end{figure}

Then, the chiral-odd DiFF $H_1^{\sphericalangle}$ was extracted from the Artru-Collins asymmetry by integrating upon the hemisphere of the antiquark jet. The experimental data for $A_{e+e-}$ are organized in a $(z,M_h)$ grid of 64 
bins~\cite{Vossen:2011fk}. They were fitted starting from an expression for $H_1^{\sphericalangle\, u}$ at $Q_0^2=1$ GeV$^2$ with 9 parameters~\cite{Courtoy:2012ry}, and then evolving it to the Belle scale. The fit and the error analysis were carried out using a Monte Carlo approach because we noticed that the minimization pushes the theoretical function towards its upper or lower bounds, where the prerequisites for the standard Hessian method are not valid. The Monte Carlo approach is inspired to the work of the NNPDF collaboration~\cite{Forte:2002fg}, although our results are not based on a neural-network fit. The approach consists in creating $N$ replicas of the data points by shifting them by a Gaussian noise with the same variance as the measurement. Each replica, therefore, represents a possible outcome of an independent measurement. Then, the standard minimization procedure is applied to each replica separately (for details, see Ref.~\cite{Bacchetta:2012ty}). The number of replicas is chosen so that the mean and standard deviation of the set of replicas accurately reproduces the original data points. In our case, it turns out to be 100. 

\begin{figure}
\centering
\includegraphics[width=7cm]{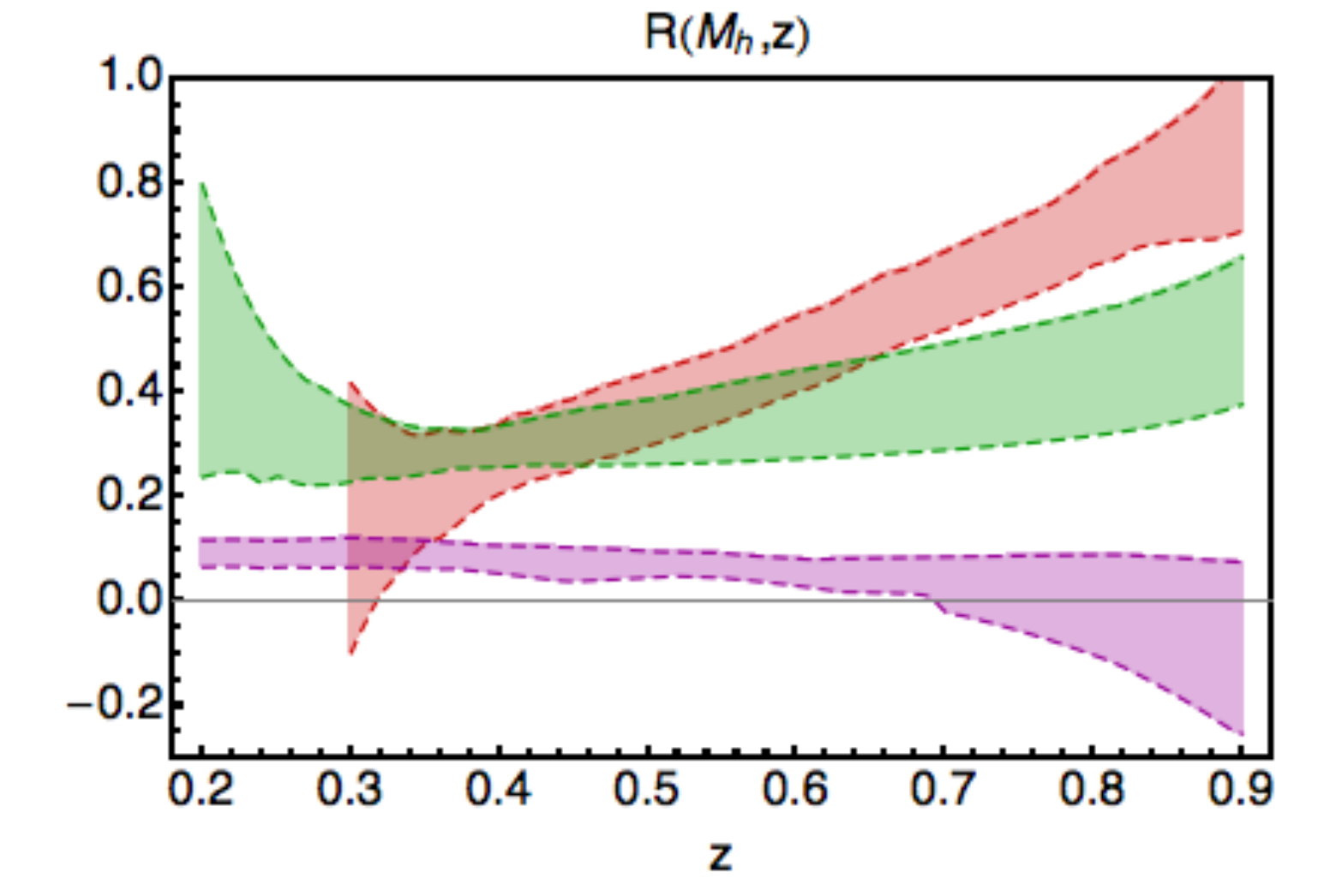}
\caption{$(|{\bf R}| / M_h) \, (H_1^{\sphericalangle\, u} / D_1^u)$ as a function of  $z$ at $Q_0^2=1$ GeV$^2$ for three different $M_h=0.4$ GeV (lower violet band), $M_h=0.8$ GeV (mid green band), and $M_h=1$ GeV (upper red band). }
\label{fig:H1z}
\end{figure}

In Fig.~\ref{fig:H1Mh}, the ratio $(|{\bf R}|/M_h) \, (H_1^{\sphericalangle\, u}/D_1^u)$ at $Q_0^2=1$ GeV$^2$ is reported as a function of $M_h$ for the three different $z=0.25$ (shortest violet band), $z=0.45$ (lower green band), and $z=0.65$ (upper red band). In Fig.~\ref{fig:H1z}, the same quantity is plotted as a function of $z$ for the three different $M_h=0.4$ GeV (lower violet band), $M_h=0.8$ GeV (mid green band), and $M_h=1$ GeV (upper red band). In both cases, the uncertainty bands correspond to the $68\%$ of all 100 replicas, produced by rejecting the largest and lowest $16\%$ among the replicas' values for each $M_h$ or $z$ point. These results should be compared with Fig.~6 of Ref.~\cite{Courtoy:2012ry}. 

%

\begin{figure}
\centering
\includegraphics[width=7cm]{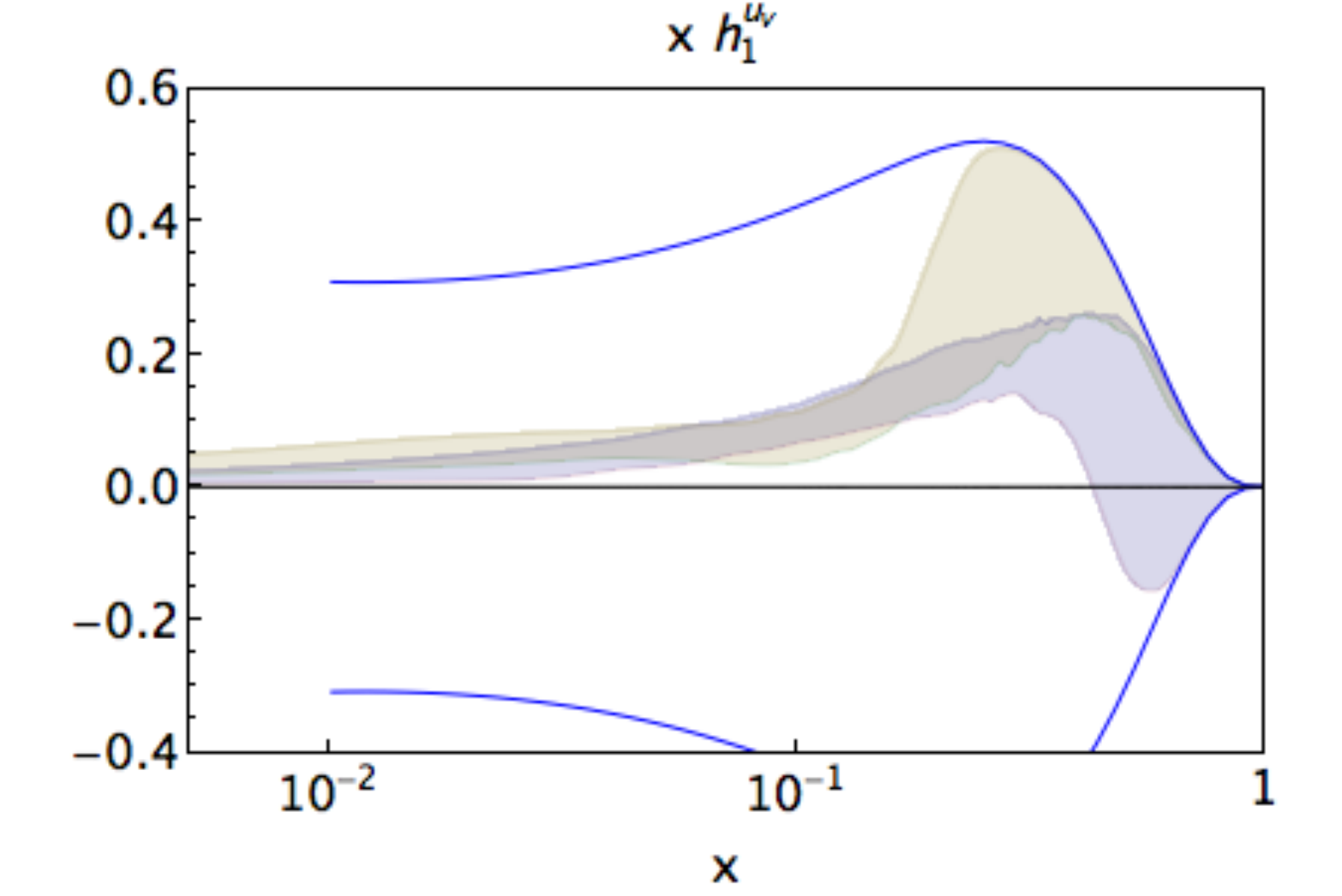}
\caption{The $x h_1^{u_v}$ as a function of $x$ at $Q^2=2.4$ GeV$^2$ in the flexible scenario. The band in the background is the result of Ref.~\cite{Bacchetta:2012ty}. The grey band in the foreground is the present new result. The blue thick solid lines indicate the Soffer bound.}
\label{fig:xh1}
\end{figure}

\section{Extraction of transversity}
\label{sec:h1}

The valence components of transversity are extracted from Eqs.~(4) 
and (\ref{eq:xh1D}) by inserting the obtained DiFFs in $n_q$ and $n^\uparrow_u$ and evolving them at each $Q^2$ of the experimental points in $x$. The unpolarized distributions $f_1^q$ are taken from the MSTW08 set~\cite{Martin:2009iq} at leading order (LO). The experimental data in $A^p_{{\rm SIDIS}}$ and $A^D_{{\rm SIDIS}}$ are also replicated 100 times by shifting them by a Gaussian noise with the same variance as the measurement, and the resulting expressions are separately fitted (for more details on the procedure, see Ref.~\cite{Bacchetta:2012ty}). For each replica of the asymmetry data, a corresponding replica of DiFFs is selected. 

Transversity must satisfy the Soffer's inequality~\cite{Soffer:1995ww} at each $Q^2$. But if it does at some initial $Q_0^2$, it will do also at higher $Q^2 \geq Q_0^2$. Thus, we impose this condition by building the following functional form at 
$Q_0^2 = 1$ GeV$^2$: 
\begin{eqnarray} 
x\, h_1^{q_v}(x; Q_0^2) &=
&\tanh \Bigl[ x^{1/2} \, \bigl( A_q+B_q\, x+ C_q\, x^2 \nonumber \\
& &\hspace{1.5cm} +D_q\, x^3\bigr)\Bigr]\nonumber \\
& &\times \; x \, 
\Bigl[ \mbox{\small SB}^q(x; Q_0^2)+ \mbox{\small SB}^{\bar q}(x; Q_0^2)\Bigr] \, ,
\label{eq:funct_form}
\end{eqnarray} 
where the analytic expression of the Soffer bound SB$^q(x; Q^2)$ can be found in Appendix of Ref.~\cite{Bacchetta:2012ty}. The hyperbolic tangent is such that the Soffer bound is always fulfilled. The low-$x$ behavior is determined by the $x^{1/2}$ term, which is imposed by hand to grant the integrability of Eq.~(\ref{eq:funct_form}) and a finite tensor charge. Present fixed-target data do not allow to constrain it. The functional form is very flexible and can contain up to three nodes. Here, we show the results employing all parameters but the $D_q$ ones, the so-called flexible scenario (for results with other choices, see Ref.~\cite{Bacchetta:2012ty}). 

In Fig.~\ref{fig:xh1}, the $q = u_v$ contribution in Eq.~(\ref{eq:funct_form}) is shown for the flexible scenario as a function of $x$ and evolved at $Q^2=2.4$ GeV$^2$. The uncertainty bands correspond again to the $68\%$ of all the 100 replicas, produced by rejecting the largest and lowest $16\%$ among the replicas' values at each $x$ point. The band in the background corresponds to our previous analysis published in Ref.~\cite{Bacchetta:2012ty}. The grey band in the foreground is the new preliminary result. The difference in the analyses is twofold. Firstly, a more realistic description of the uncertainties on the extraction of DiFFs is obtained by using the replica method, as already explained in the previous section. Then, the most recent and more precise data for $A^p_{{\rm SIDIS}}$ from COMPASS are used~\cite{Adolph:2014fjw}. The combined effect produces a narrower band than before where there are experimental data, {\it i.e.} for the interval $0.0064 \leq x \leq 0.28$. For larger $x$, the replicas tend to fill all the available phase space within the Soffer bound, that is represented by the two blue solid lines. The replicas graphically visualize the realistic degree of uncertainty on the knowledge of transversity, in particular where there are no experimental data. It is interesting to note that the new result shows a band which is on average smaller than the previous one. Since the latter was in overall agreement with the transversity extracted at the same scale through the Collins effect~\cite{Anselmino:2013vqa}, it seems that altering this agreement poses a serious question about what is the real size of transversity. 

Since the data for $A^D_{{\rm SIDIS}}$ are the same as the previous analysis, nothing changes in the result for the valence down component of transversity. In particular, the agreement with the $h_1^{d_v} (x)$ extracted through the Collins effect is good except for $x \geq 0.1$~\cite{Bacchetta:2012ty}. In this range, our results tend to saturate the lower limit of the Soffer bound because they are driven by the COMPASS deuteron data, in particular by the bins number 7 and 8. This statement remains valid also for the other scenarios, indicating that this is not an artifact of the chosen functional form. 

\begin{figure}
\centering
\includegraphics[width=7cm]{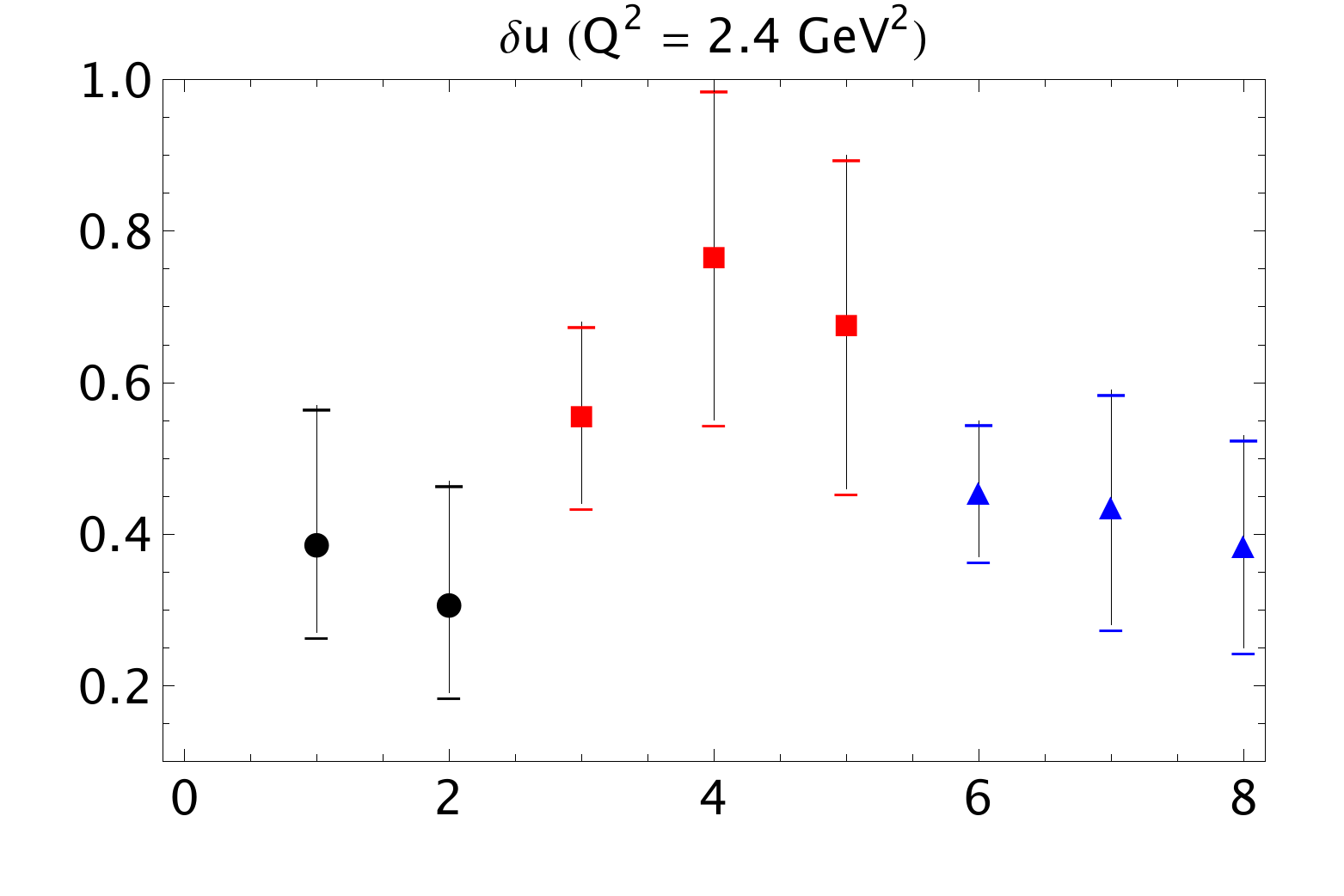}
\caption{Tensor charges for the valence up quark. From left to right: black filled circles (numbers 1 and 2) for the two values obtained through the Collins effect in Ref.~\cite{Anselmino:2008jk}, red squares (numbers 3-5) for the three values obtained in the previous analysis of Ref.~\cite{Bacchetta:2012ty} using three different fitting forms, blue triangles (numbers 6-8) for the corresponding three values obtained in the present new analysis.}
\label{fig:tensor}
\end{figure}

In Fig.~\ref{fig:tensor}, the first moment in $x$ of the valence up component of transversity is displayed at $Q^2=2.4$ GeV$^2$, namely the tensor charge $\delta u_v$ for the valence up quark in the proton. The integration range is extended to the lower $(x=0)$ and upper $(x=1)$ limits by extrapolating $h_1^{u_v} (x)$ outside the $x$ range of experimental data. The two leftmost black filled circles (numbers 1 and 2) are the results obtained from the analysis of the Collins effect using two different methods for the extraction of the Collins function from $e^+ e^-$ annihilation data~\cite{Anselmino:2008jk}. The next set of three red squares (numbers 3 to 5) corresponds to the results of the previous analysis in Ref.~\cite{Bacchetta:2012ty} when one, two, or three nodes are considered in the fitting functional form of Eq.~(\ref{eq:funct_form}). The three rightmost blue triangles (numbers 6 to 8) are the results of the present new analysis; the flexible scenario here considered corresponds to the central triangle (number 7). As a consequence of results displayed in the previous figure, the value of the tensor charge is smaller than in the previous analysis, but the large uncertainty related to extrapolation makes almost all of the obtained values consistent. Again, since the experimental data for $A^D_{{\rm SIDIS}}$ are the same as before, also the values for the valence down tensor charges are basically unaltered. The numerical values of the tensor charge obtained in the flexible scenario for both up and down valence quarks are 
\begin{equation}
\delta u_v = 0.44 \pm  0.15  \quad \delta d_v = -0.35 \pm  0.45  \: . 
\label{eq:tensorvalues}
\end{equation}
Finally, it must be remarked that the above results are heavily influenced by the adopted functional form, in particular by the $x^{1/2}$ power behavior in Eq.~(\ref{eq:funct_form}), which reflects the uncertainty because of missing data at very low $x$.

\section{Concluding remarks}
\label{sec:end}

Semi-inclusive production of hadron pairs is a useful tool for extracting the transversity distribution in the framework of collinear factorization by combining data for SIDIS and $e^+ e^-$ annihilation processes. In this framework, the expression for the SIDIS single spin asymmetry contains a simple product of the transversity and of its chiral-odd partner describing the fragmentation of a transversely polarized quark into an unpolarized hadron pair (Dihadron Fragmentation Function, DiFF). The latter can be extracted from a similarly simple azimuthal asymmetry for $e^+ e^-$ annihilation into two hadron pairs. Evolution equations are known to connect DiFFs at different scales. 

In this work, we firstly have updated the extraction of DiFFs in Ref.~\cite{Courtoy:2012ry} by re-performing the fit using the replica method, which allows for a more realistic estimate of the uncertainty~\cite{Bacchetta:2012ty}. 

Then, we have updated the extraction of the valence components of transversity published in Ref.~\cite{Bacchetta:2012ty} by using the previous result, and by employing the most recent and more precise SIDIS data for unidentified hadron pairs on a proton target by COMPASS~\cite{Adolph:2014fjw}. Thus, the (preliminary) results displayed in this contribution represent the current most realistic estimate of the uncertainties on our knowledge of transversity. 

According to this new analysis, the valence up component of transversity has a narrower uncertainty band than before and it turns out smaller on average. This result poses a serious question about what is the real size of transversity, since the previous overall agreement between the results of Ref.~\cite{Bacchetta:2012ty} and the ones obtained through the Collins 
effect~\cite{Anselmino:2013vqa} seems deteriorated. 

More studies are of course needed. In particular, further developments point towards the need for data on unpolarized cross sections for hadron pair production, in order to extract from data also the unpolarized DiFF (which is currently estimated using a Monte Carlo simulation). Moreover, a better error analysis requires also to go beyond the bias introduced by the choice of the fitting functional form for transversity. This task will probably be accomplished by adopting an error analysis based on Neural Networks.




%
\bibliography{mybiblio}
%
%
%
%

\end{document}